\documentclass[prb,showpacs,preprintnumbers,amsmath,amssymb,twocolumn,superscriptaddress]{revtex4-1}

\usepackage[dvipdfmx]{graphicx}
\usepackage{dcolumn}
\usepackage{bm}
\usepackage{color}

\usepackage{amsmath}	
\usepackage{amsfonts}

\newcommand{\vecc}[1]{\mbox{\boldmath $#1$}}

\begin{document}

\title{
Cyclotron resonance in Kondo insulator
}

\author{Yasuhiro Tada}
\email[]{tada@issp.u-tokyo.ac.jp}
\affiliation{Institute for Solid State Physics, University of Tokyo, Kashiwa 277-8581, Japan}

\begin{abstract}
Motivated by the recent experimental 
observations of quantum oscillations in the Kondo insulators SmB$_6$ and YbB$_{12}$,
we study magneto-optical conductivity of Kondo insulators. 
We show that there exist non-trivial cyclotron resonance peaks in a frequency domain below the hybridization gap 
as a result of the characteristic Landau level structure of the system.
Furthermore, 
these low energy peaks are enhanced near the insulator-metal 
phase transition driven by an applied magnetic field 
and stable even at low temperatures. 
They are smoothly connected via Kondo effect 
to resonance peaks for unhybridized conduction and $f$-electrons
at high temperature.
The non-trivial cyclotron resonance is a dynamical counterpart of quantum oscillations in static quantities,
and through this intimate connection it will provide a useful insight for the latter.
\end{abstract}


\maketitle

\section{Introduction}
Metals and insulators are distinguished by their low energy behaviors
which are normally well characterized by existence or absence of a Fermi surface.
Especially, their difference becomes significant under a magnetic field, where the Fermi surface is 
discretized into Landau levels and quantum oscillations can be seen as a direct consequence in metals,
while such behaviors are absent in insulators according to the standard theories~\cite{QO1984}.
However, these canonical understandings of metals and insulators are challenged by
the recent experimental observations of quantum oscillations in the Kondo insulators
SmB$_6$ and YbB$_{12}$
~\cite{Li2014,Tan2015,Xiang2018,Liu2018}.
Quantum oscillations are observed in the magnetization of SmB$_6$
~\cite{Li2014,Tan2015}, 
and the system becomes metallic for very strong magnetic fields $B\gtrsim100$T
~\cite{Cooley1999,Chen2015,Shahrokhvand2017,Wolgast2017}.
In YbB$_{12}$, quantum oscillations are found when $B\sim 40$T~\cite{Xiang2018,Liu2018},
and a field-induced metallic phase is realized for $B\gtrsim 50$T
~\cite{Sugiyama1988,Terashima2017,Terashima2018}.
In this way,
both of these materials show quantum oscillations inside the insulating phases and they are enhanced 
as the system approaches the
insulator-metal phase transition point.
In addition, SmB$_6$ and YbB$_{12}$ are candidates of topological Kondo insulators~\cite{TKI2010,TKI2016},
and possible relations of their topological nature to the quantum oscillations attract great attention.

To understand these anomalous quantum oscillations, a number of theories have been proposed 
~\cite{KnolleCooper2015,Zhang2016,Pal2016,Ram2017,Shen2018,Grubinskas2018,Peters2019,
Erten2016,KnolleCooper2017ex,Sodemann2018}.
Although theoretical understandings have not yet settled,
it was demonstrated that 
quantum oscillations can be found not only in Kondo insulators but also in general insulators
with band inversion, 
because Landau levels change their character between conduction-electron-like one and $f$-electron-like one
successively as the magnetic field increases~\cite{KnolleCooper2015,Zhang2016,Pal2016,
Ram2017,Shen2018,Grubinskas2018,Peters2019}.
This is a hallmark of the characteristic Landau level structure realized in a Kondo insulator which 
is totally different from that in a conventional semiconductor 
or a simple insulator without band inversion.
Thus far,
unusual Landau levels and anomalous properties under magnetic fields
have been extensively studied in exotic semimetals such as graphene and topological materials
~\cite{graphene_RMP,Ando2013,semimetal2018}.
Now band inverted insulators provide a new platform for physics under strong magnetic fields.
In order to pin down the origin of the observed quantum oscillations
and elucidate impacts of magnetic fields, 
it is desirable to clarify the Landau level structures and their physical consequences in detail.
However, the previous theoretical studies focus only on static quantities,
while a dynamical quantity may provide rich information for clarification of the unusual Landau level structures
and also for a possible origin of the observed quantum oscillations.

In this study,
we investigate cyclotron resonance in Kondo insulators for the first time.
The cyclotron resonance can directly access a Landau level structure 
by frequency dependence of magneto-optical conductivity,
where
a resonance peak corresponds to a transition between different Landau levels. 
Historically, it was first developed mainly for semiconductors and 
later employed for studying semimetallic graphene with unusual Landau levels
~\cite{graphene_RMP,graphene_ex2006,graphene_ex2007,graphene_ex2007R,graphene_th1998,graphene_th2007}.
Recently it was also applied to metals
with strong correlations such as the candidate topological superconductor Sr$_2$RuO$_4$,
Fe-based superconductor KFe$_2$As$_2$, and hidden order phase of the $f$-electron material URu$_2$Si$_2$
~\cite{SrRuO2000,KFeAs2011,URuSi2012}.
Similarly to the quantum oscillations in static quantities, 
it has been naively expected that low energy cyclotron resonance can be obtained only for a system
with conducting carriers, while a resonance peak could be seen only for an energy scale above a gap
in insulators.
Here, we demonstrate that there exist non-trivial cyclotron resonance peaks well below 
an activation gap in a Kondo insulator at low temperature.
Furthermore, the low energy resonance peaks are enhanced near the insulator-metal
quantum phase transition point. 
These anomalous behaviors are a direct consequence of the characteristic Landau
level structures which are responsible also for quantum oscillations. 
This is a dynamical counterpart of 
quantum oscillations in static quantities, and through this intimate connection it will provide a clue 
for the mechanism of the latter.

\section{Model and calculations}
\subsection{Hamiltonian and its spectrum}
We consider a simple model of a Kondo insulator composed of single orbital conduction electrons
and $f$-electrons with pseudo-spins in two dimensions
~\cite{KnolleCooper2015,Zhang2016,Pal2016,Ram2017,Shen2018,Grubinskas2018,Peters2019,
Erten2016,KnolleCooper2017ex,Sodemann2018}. Effects of three dimensionality will be discussed later.
Since our primary purpose is to clarify the cyclotron resonance in a general Kondo insulator,
we consider both of a non-topological insulator and topological insulator on an equal footing.
The Hamiltonian is $H=\int d^2x \psi^{\dagger}(\vecc{x}){\mathcal H}\psi(\vecc{x})$,
\begin{align}
{\mathcal H}&=\left(
\begin{array}{cc}
\frac{\pi^2}{2m_c}+\frac{g_c\mu_B}{2}\sigma_z B & v_0+v_1a\vecc{\pi}\cdot \vecc{\sigma}^t \\
v_0+v_1a\vecc{\pi}\cdot \vecc{\sigma}^t & -\frac{\pi^2}{2m_f}+\epsilon_{f0}+\frac{g_f\mu_B}{2}\sigma_zB
\end{array}
\right),
\label{eq:H}
\end{align}
where $\psi(\vecc{x})=(\psi_{c\uparrow},\psi_{c\downarrow},\psi_{f\uparrow},\psi_{f\downarrow})$
is the field operator for the conduction and $f$-electrons.
$\vecc{\pi}=-i\hbar\nabla+e\vecc{A}$ with the electric charge $-e<0$ is the dynamical momentum
and $\vecc{A}$ is a vector potential 
for a uniform magnetic field $\vecc{B}=\nabla\times \vecc{A}=(0,0,B)$ with $B>0$.
$m_c, m_f$ are effective masses for the conduction electrons and $f$-electrons respectively,
and $m_c$ is assumed to equal the bare electron mass $m_0$ and 
$m_f=0.1m_c$ for brevity.
$\epsilon_{f0}$ is the $f$-electron energy level, and $v_0, v_1$ represent hybridization strength 
with a characteristic lattice length scale $a\sim$ few {\AA}  which is taken to be the length unit.
The system is a topological Kondo insulator when $v_0=0, v_1\neq0$, 
while it is a non-topological one for $v_0\neq0, v_1=0$~\cite{TKI2010,TKI2016}.
$\mu_B=e\hbar/2m_0$ is the Bohr magneton and $g$-factors are assumed to be $g_c=g_f=2$ for simplicity.
We use the energy unit $t_0\equiv \hbar^2/(2m_0a^2)$, and 
set $\hbar =1$ and the Boltzmann constant $k_B=1$.
We fix the model parameters as $\epsilon_{f0}=0.5t_0$, and
$(v_0,v_1)=(0.05t_0,0)$ for a non-topological insulator and $(v_0,v_1)=(0,0.05t_0)$ for a topological insulator,
where the activation gaps at zero magnetic field are 
$\Delta_0=2v_0=0.1t_0$ and $\Delta_1=\sqrt{\epsilon_{f0}^2-(4(v_1a)^2m_cm_f/(m_c+m_f))^2}\simeq 0.067t_0$,
respectively.
Although these parameters may be too large when considering the Kondo insulators
SmB$_6$ and YbB$_{12}$,
they will be renormalized by the strong Coulomb interactions between 
the $f$-electrons as $m_f\rightarrow m_f/z, v_{0,1}\rightarrow\sqrt{z}v_{0,1}$ with
a renormalization factor $z\ll 1$.
We have confirmed that our results are qualitatively unchanged when we use different parameter values.

\begin{figure}[htbp]
\begin{tabular}{cc}
\begin{minipage}{0.5\hsize}
\begin{center}
\includegraphics[width=\hsize]{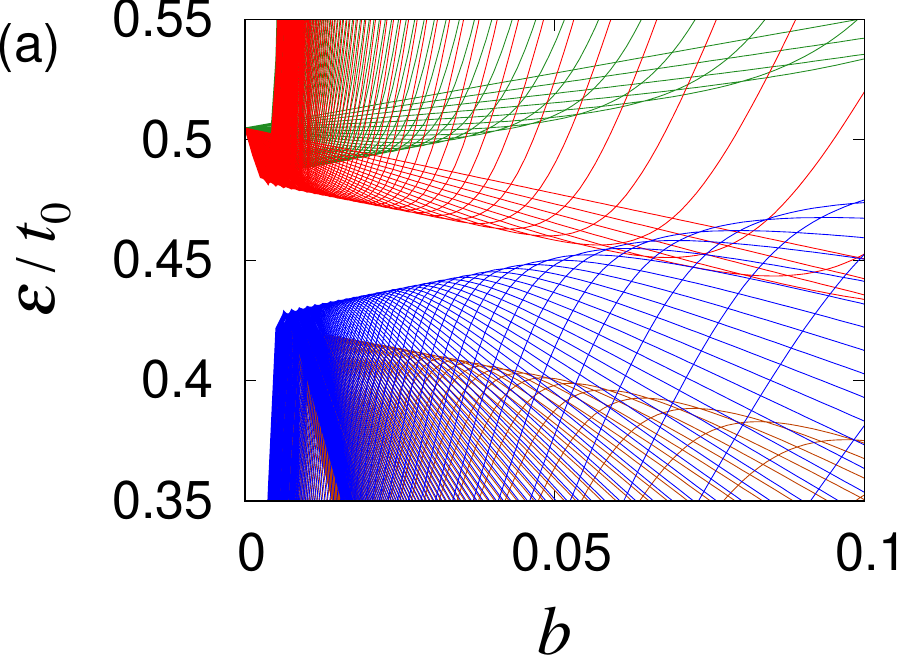}
\end{center}
\end{minipage}
\begin{minipage}{0.5\hsize}
\begin{center}
\includegraphics[width=\hsize]{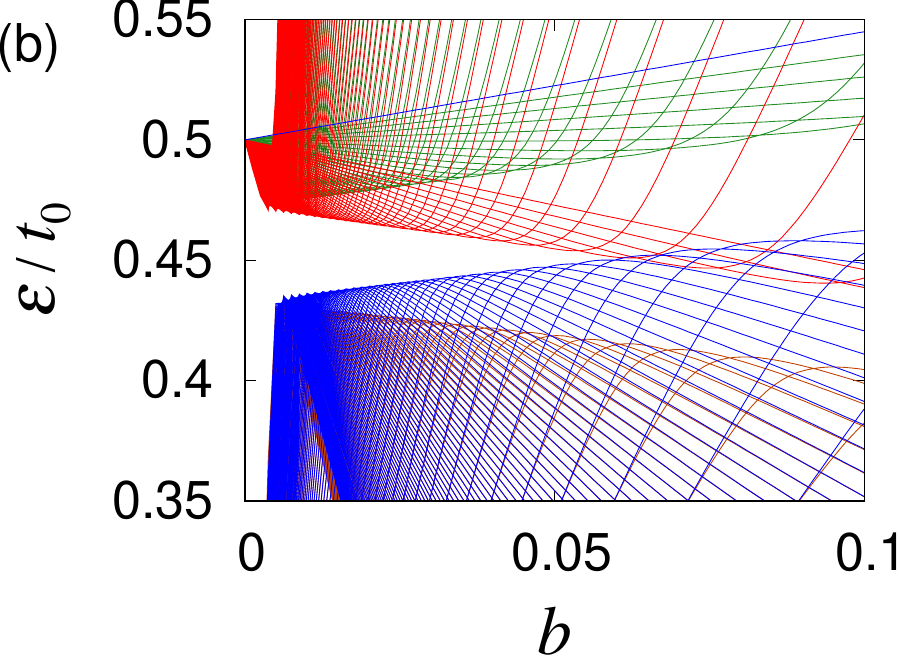}
\end{center}
\end{minipage}
\end{tabular}
\caption{Energy spectra of (a) the non-topological and (b) topological Kondo insulators.
The dimensionless magnetic field is $b= eB/(m_0t_0)
=4\pi Ba^2/\phi_0$.
The different colors correspond to the four eigenvalues $\{ \varepsilon_{nm}\}_{m=1}^4$ for each Landau level $n$.}
\label{fig:energy}
\end{figure}

It is straightforward to diagonalize the Hamiltonian \eqref{eq:H}.
We expand the field operator as $\psi_{i}(\vecc{x})=\sum_{nk}a_{nki}\phi_{nk}(\vecc{x})$ 
($i$ represents $i=(c,\uparrow)$ etc.)
by using the harmonic oscillator single-particle wave functions 
$\phi_{nk}$ with a Landau level $n$ and an index for degeneracy $k$.
$a_{nk}$ is the corresponding annihilation operator.
For a system in the infinite cylinder geometry with the size $L_x=\infty, L_y<\infty$,
the lowest Landau level wave function is $\phi_{0k}=1/\sqrt{\sqrt{\pi}l_BL_y}\exp(iky-(x-kl_B^2)^2/2l_B^2)$ 
under the Landau gauge $\vecc{A}=(0,Bx,0)$. 
They satisfy $l_B\pi_+\phi_{nk}=\sqrt{n+1}\phi_{n+1k}, l_B\pi_-\phi_{nk}=\sqrt{n}\phi_{n-1k}$, 
where $\pi_{\pm}=(\pi_x\pm i\pi_y)/\sqrt{2}$ and $l_B=1/\sqrt{eB}$ is the magnetic length.
We suppress the index $k$ because it is not directly involved in the following discussions. 
The Hamiltonian ${\mathcal H}_{n,ij}$ in the Landau level basis gives eigenvalues $\{\varepsilon_{nm}\}_{m=1}^4$
and eigenvectors $\{u_{n,im}\}_{m=1}^4$\cite{KnolleCooper2015,Zhang2016}.
The energy spectra for both the non-topological and topological cases are shown in Fig. \ref{fig:energy}.
The gap closes at a critical value $b_c\simeq 0.058$ with $b\equiv eB/(m_0t_0)
=4\pi Ba^2/\phi_0$ ($\phi_0=h/e$ is the magnetic flux quantum) for both $(v_0,v_1)=(0.05t_0,0), (v_0,v_1)=(0,0.05t_0)$
and the system becomes metallic for $b>b_c$,
corresponding to a field induced metallic phase as seen in
SmB$_6$ and YbB$_{12}$
~\cite{Cooley1999,Chen2015,Shahrokhvand2017,Wolgast2017,Sugiyama1988,Terashima2017,Terashima2018}.
For simplicity, we set the chemical potential so that it coincides with the energy where the gap closes,
$\mu=0.454t_0$ for the non-topological case and $\mu=0.45t_0$ for the topological case, respectively.
The Landau level structure is complicated, but roughly speaking, 
it is a combination of $\varepsilon_{cn}=\omega_c(n+1/2)$ and $
\varepsilon_{fn}=-\omega_f(n+1/2)+\epsilon_{f0}$ with Zeeman shifts,
and Landau levels change their characters between conduction-electron-like one and $f$-electron-like one
successively as $b$ increases.
Because of the hybridization $v_{0,1}$,
energy spacing between the neighboring Landau levels
is not exactly $\omega_c, \omega_f$ but can be in between them especially when 
the crossover in Landau level's character takes place,
which is responsible for quantum oscillations in static qunatities.
We will see in the following that the magneto-optical conductivity can faithfully
reflect these characteristic structures, providing rich information on the unusual Landau levels.

\subsection{Brief overview of dHvA effect}
\begin{figure}[htbp]
\includegraphics[width=0.6\hsize]{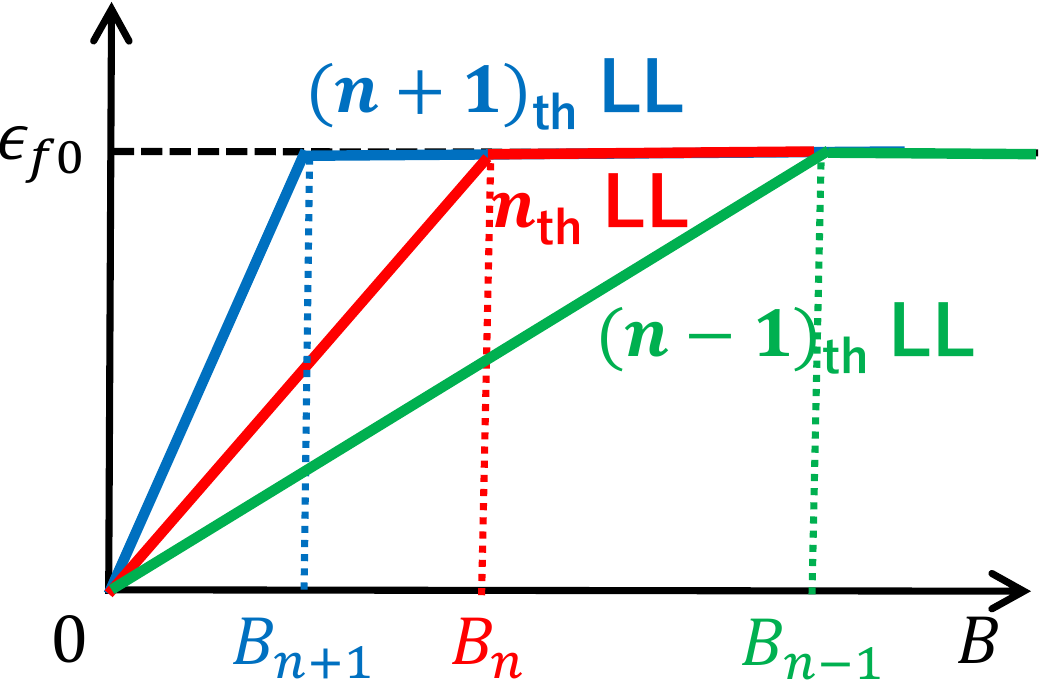} 
\caption{The approximte Landau level (LL) spectrum $\{\varepsilon_{n-}\}$ (Eq.~\eqref{eq:e_approx}) 
and the crossover fields $\{B_n\}$.
Only three Landau levels are shown for simplicity. 
}
\label{fig:e_}
\end{figure}

Before going to the discussion on cyclotron resonance,
we first give a brief overview on the quantum oscillations of orbital magnetization
at zero temperature (de Haas-van Alphen (dHvA)
effect) to provide an intuitive insight of the characteristic Landau levels
~\cite{KnolleCooper2015}. 
This will be helpful for understanding the intimate relationship between the quantum oscillations
and cyclotron resonance discussed in the next section.
In this section, we keep the Planck constant $\hbar$ explicitely.
For simplicity we ignore the Zeeman effect and consider $m_f\rightarrow\infty, v_0\neq0, v_1=0$
(a similar discussion holds for the topological $v_0=0, v_1\neq0$ case).
In this case, the energy spectrum is given by 
\begin{align}
\varepsilon_{n\pm}=\frac{1}{2}[\varepsilon_{cn}+\epsilon_{f0}\pm 
\sqrt{(\varepsilon_{cn}-\epsilon_{f0})^2+4v_0^2}],
\label{eq:e_}
\end{align}
where $\varepsilon_{cn}=\hbar\omega_c(n+1/2)$.
At zero temperature limit $T\rightarrow 0$, only the lower energies $\varepsilon_{n-}<\mu$ contribute
to the ground state energy.
In the limit of small hybridization $v_0\ll \epsilon_{f0}$, 
it behaves as $\varepsilon_{n-}\simeq \varepsilon_{nc}-(\varepsilon_{cn}^2+4v_0^2)/\epsilon_{f0}$ for 
$B\ll B_n$, while $\varepsilon_{n-}\simeq \epsilon_{f0}-(\epsilon_{f0}^2+4v_0^2)/\varepsilon_{cn}$ for 
$B\gg B_n$, where $B_n$ is the crossover scale between the conduction-electron-like character 
and $f$-electron-like character defined by $\hbar\omega_c(n+1/2)=\epsilon_{f0}$,
or equivalently $B_n=m_c\epsilon_{f0}/(e\hbar(n+1/2))$.
To perform a simple analytical calculation, we introduce an approximation where
$\varepsilon_{n-}$ is replaced by
\begin{align}
\varepsilon_{n-}=\left\{
\begin{array}{ll}
\varepsilon_{cn} & (B\leq B_n) \\
\epsilon_{f0} & (B>B_n)
\end{array}
\right..
\label{eq:e_approx}
\end{align}
The schematic behavior of $\varepsilon_{n-}$ is shown in Fig.~\ref{fig:e_}.
When the magnetic field is $B_{n+1}<B<B_n$, the ground state energy density is 
\begin{align}
u(B)&=D\sum_{l=0}^{n}\varepsilon_{l-}+\tilde{u}_{}\nonumber\\
&=D\frac{\hbar\omega_c}{2}[n(n+1)+n+1]+\tilde{u}_{},
\end{align}
where $D=2/(2\pi l_B^2)=eB/\pi\hbar$ is the degeneracy per area including spin degrees of freedom
and $\tilde{u}_{}=D\sum_{l>n}\varepsilon_{l-}=D\sum_{l>n}\epsilon_{f0}$ 
is the contribution from higher Landau levels $n+1,n+2,\cdots$
with an appropriate regularization to avoid unphysical divergence.
Similarly, the ground state energy density for $B_n<B<B_{n-1}$ is 
\begin{align}
u(B)&=D\frac{\hbar\omega_c}{2}[(n-1)n+n]+D\epsilon_{f0}+\tilde{u}_{},
\end{align}
where the second term is the contribution from $\varepsilon_{n-}$ and the 
functional form of $\tilde{u}$ is the same as above.
Correspondingly, the orbital magnetization $M=-\partial u/\partial B$ is given by
\begin{align}
M(B_{n+1}<B<B_n)
&=-\frac{e\omega_c}{\pi}[n(n+1)+n+1]+\tilde{M},\\
M(B_{n}<B<B_{n-1})
&=-\frac{e\omega_c}{\pi}[(n-1)n+n]-\frac{e\epsilon_{f0}}{\pi\hbar}+\tilde{M},
\end{align}
where $\tilde{M}(B)=-\partial \tilde{u}_{}/\partial B$ and does not have a singularity.
From these expressions, we see that the magnetization jumps at $B=B_n$
and oscillates with the period $\delta(1/B)=1/B_n-1/B_{n-1}=e\hbar/(m_c\epsilon_{f0})$
which is inversely proportional to the area of a Fermi surface cross section
without hybridization when $\epsilon_{f0}\simeq\mu$.
The size of the jump gives the oscillation amplitude, 
$\delta M(B=B_n)=M(B_n+0)-M(B_n-0)= e\epsilon_{f0}/(\pi\hbar)$.
Note that
the sharp discontinuity of $M(B)$ is due to the approximation Eq.~\eqref{eq:e_approx} where
$\varepsilon_{n-}$ is not differentiable at $B=B_n$, 
and $M(B)$ is smooth for the original spectrum Eq.~\eqref{eq:e_}.
Note also that the condition $m_f\rightarrow\infty$ can be relaxed and the quantum oscillation can be
described by a similar argument for $m_f<\infty$.

From the above derivation of the oscillating magnetization $M(B)$ where all the states below the
chemical potential contribute,
it is clear that the crossover of a Landau level character as a function of $B$
from conduction-electron-like one to $f$-electron-like one
is the most important key.
This observation holds true also for the quantum oscillation for static transport (Shubnikov-de Haas effect)
where an effective density of states thermally smeared around the chemical potential  
will play a central role~\cite{Zhang2016}. 
The smeared density of states oscillates also due to the crossover of the Landau level character
as easily expected from Fig.~\ref{fig:energy} where a Landau level energy approaches the band edge
and goes away from it as $B$ increases.
Therefore, we conclude that the quantum oscillations in static quantities are essentially
due to the Landau level crossover in the present model.
Note that there are densely distributed Landau levels near the edges of upper/lower bands 
in Fig.~\ref{fig:energy},
because energy level spacings becomes narrower when the slope of a Landau level energy changes
around the crossover field strength.   
In the next section, we will see that this characteristic structure of the Landau levels leads to
low energy cyclotron resonance, sharing essentially same origin with the quantum oscillations of
static quantities.
.

\subsection{Cyclotron resonance}

We calculate the magneto-optical conductivity for a circularly polarized light with frequency $\omega$
~\cite{Mahan2000},
\begin{align}
\textrm{Re}\sigma_{-+}(\omega)&=\frac{1}{\omega}\left(\textrm{Im}K_{-+}^R(\omega)
-\textrm{Im}K_{-+}^R(0)\right),
\label{eq:sigma}\\
K_{-+}^R(\omega)&=\frac{-i}{V}\int dt\theta(t)\langle [J_-(t),J_+(0)]\rangle e^{i\omega t}\nonumber\\
&=\sum_{nmm'}M^{-+}_{n,mm'}\chi^{-+}_{n,mm'}(\omega),
\label{eq:K}\\
\chi^{-+}_{n,mm'}(\omega)&=\frac{f(\varepsilon_{n+1,m})-f(\varepsilon_{nm'})}
{\omega -\varepsilon_{n+1,m}+\varepsilon_{nm'}+i\gamma},
\label{eq:chi}
\end{align}
where $V$ is the volume of the system and $f(\varepsilon)=(e^{(\varepsilon-\mu)/T}+1)^{-1}$.
Similarly, the conductivity for the opposite polarization $\sigma_{+-}(\omega)$ is given by the correlation function
$K^R_{+-}(\omega)=\frac{-i}{V}\int dt\theta(t)\langle [J_+(t),J_-(0)]\rangle e^{i\omega t}$,
and satisfies $\textrm{Re}\sigma_{-+}(\omega)=\textrm{Re}\sigma_{+-}(-\omega)$.
The current operators are given by 
$J_{+}=J_-^{\dagger}=J_x+ iJ_y
=\sum_{}{\mathcal J}_{+n,ij}a^{\dagger}_{n+1,i}a_{nj}$ with
$J_{x,y}=\int d^2x \psi^{\dagger} \partial{\mathcal H}/\partial A_{x,y}\psi$,
and ${\mathcal J}_{n+}={\mathcal J}_{n-}^{\dagger}$ is defined by 
\begin{align}
{\mathcal J}_{n+}&=
-\sqrt{2}e\left(
\begin{array}{cccc}
\frac{\sqrt{n+1}}{m_cl_B} & 0 & 0 & \sqrt{2}v_1a\\
0 & \frac{\sqrt{n}}{m_cl_B} & 0 & 0 \\
0 & \sqrt{2}v_1a &  -\frac{\sqrt{n+1}}{m_fl_B} & 0\\
0 & 0 & 0 & -\frac{\sqrt{n}}{m_fl_B}
\end{array}
\right).
\end{align}
The matrix $M$ is defined as $M_{n,mm'}^{-+}=
\textrm{tr}[{\mathcal J}_{n-}{\mathcal U}_{n+1m}{\mathcal J}_{n+}{\mathcal U}_{nm'}]/(2\pi l_B^2)$ with
$({\mathcal U}_{nm})_{ij}=u_{n,im}u_{n,jm}^{\ast}$.
We fix the phenomenological broadening parameter $\gamma =0.0005t_0$ and results are essentially
independent of $\gamma$.

\begin{figure}[htbp]
\begin{tabular}{cc}
\begin{minipage}{0.5\hsize}
\begin{center}
\includegraphics[width=\hsize]{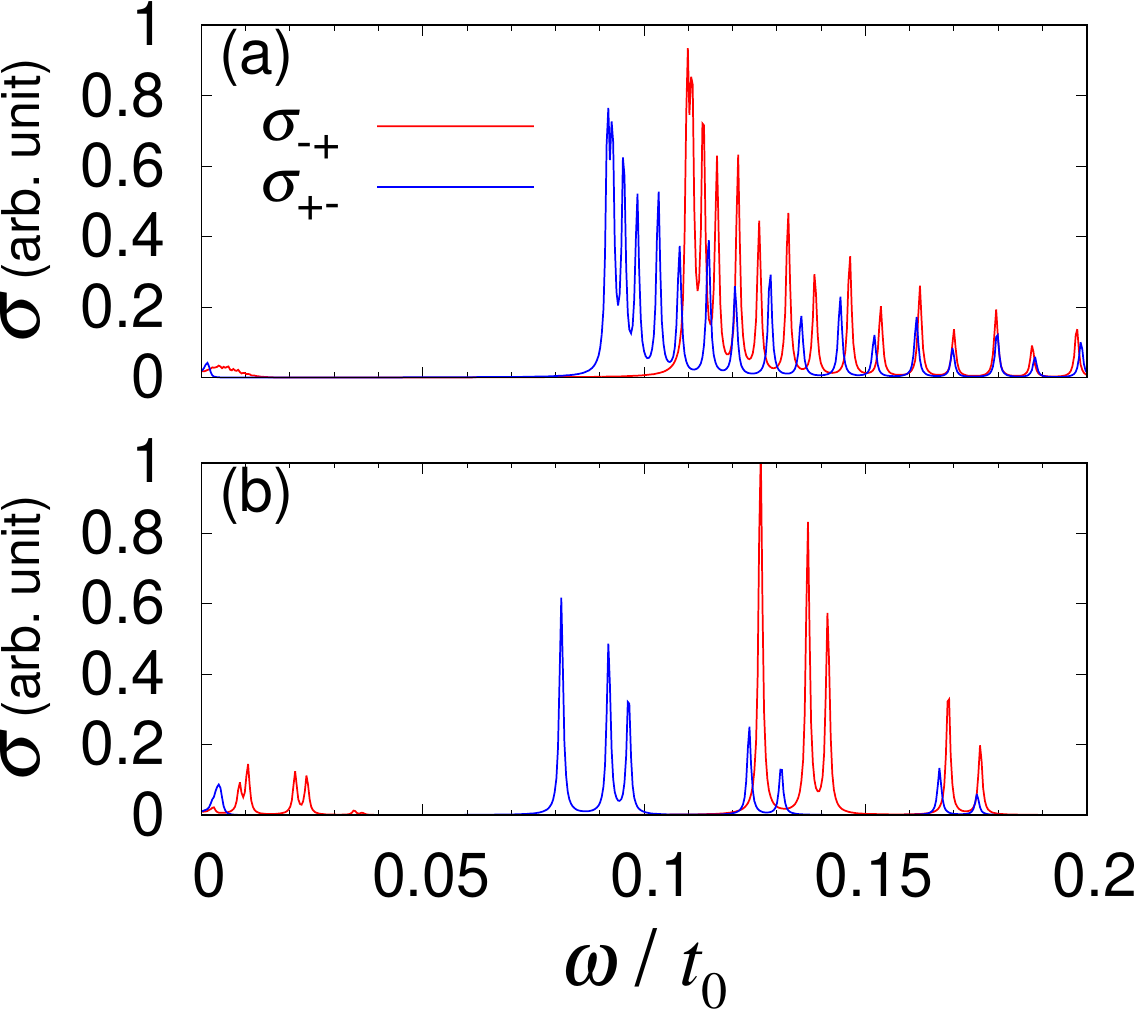}
\end{center}
\end{minipage}
\begin{minipage}{0.5\hsize}
\begin{center}
\includegraphics[width=\hsize]{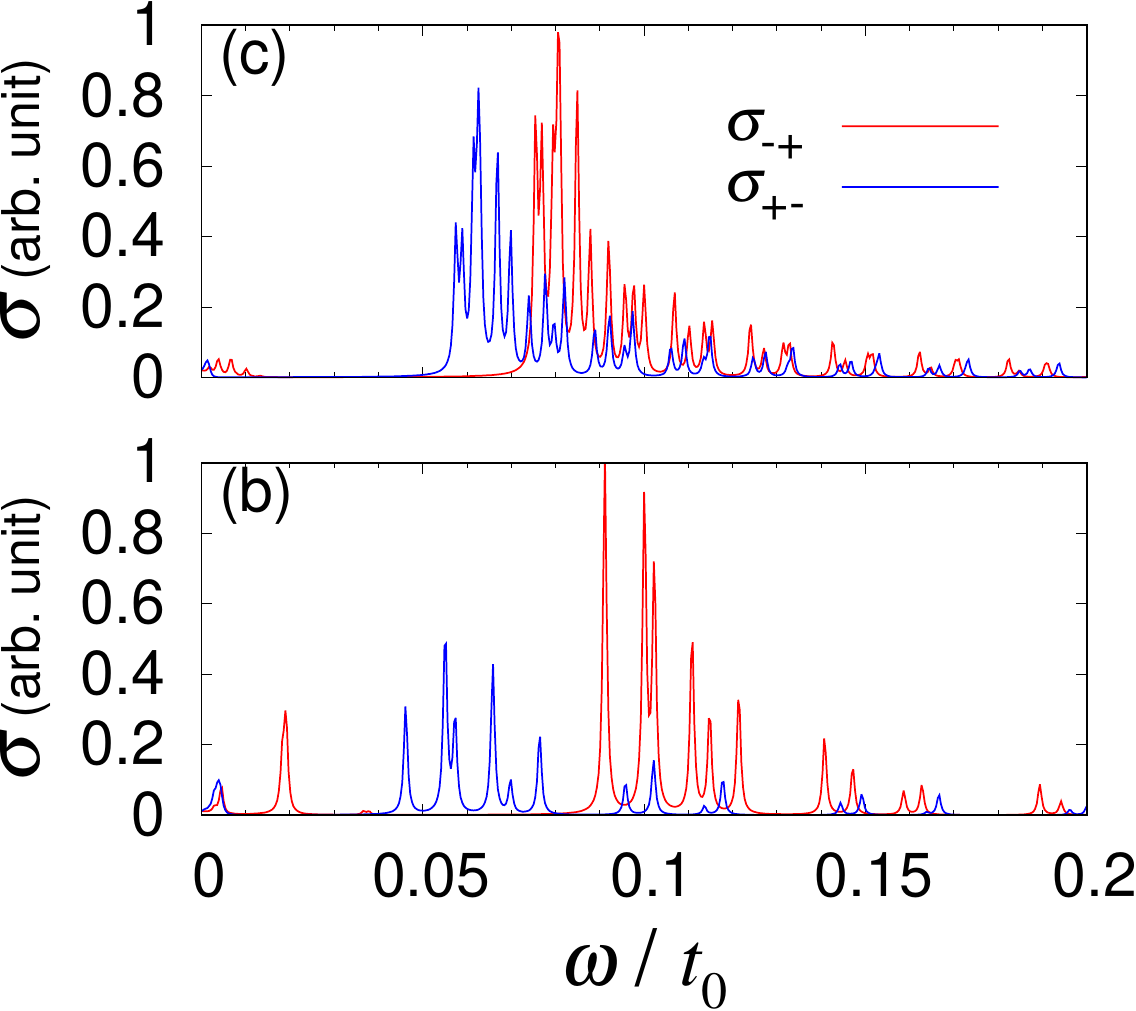}
\end{center}
\end{minipage}
\end{tabular}
\caption{Real part of the magneto-optical conductivity (in arbitrary unit) for 
(a), (b) the non-topological insulator at $T=0.007t_0= 0.07\Delta_0$ and 
(c), (d) the topological insulator at $T=0.0047t_0\simeq 0.07\Delta_1$.
The magnetic field is $b=0.02\simeq0.33b_c$ for (a), (c)  and $b=0.05\simeq0.83b_c$ for (b), (d).
The red curve is Re$\sigma_{-+}(\omega)$ and the blue one is
Re$\sigma_{+-}(\omega)$. 
}
\label{fig:om}
\end{figure}

Figure \ref{fig:om} shows the magneto-optical conductivity $\sigma(\omega)$ at a low temperature $T=0.007t_0
=0.07\Delta_0$ for the non-topological case  
and $T=0.0047t_0\simeq 0.07\Delta_1$ for the topological case.
The calculated cyclotron resonance is qualitatively similar for these two cases, although there are some quantitative differences.
For the small magnetic field $b=0.02\simeq 0.33b_c$, the conductivity is strongly suppressed 
and there are only small spectral weights 
in a low energy region below the activation gap $\omega<\Delta_{0,1}$. 
At high energy,
a series of large peaks are located for $\omega\gtrsim \Delta_{0,1}$. 
On the other hand,
when the magnetic field is close to the critical value, $b=0.05\simeq 0.83b_c$,
$\sigma(\omega)$ at low energy is enhanced and
there exist clear peaks well below the gap $\Delta_{0,1}$
even in the insulating states.
More precisely, the low energy peaks are located in the region $\omega_f=0.005t_0\leq \omega
\leq\omega_c=0.05t_0$ and will be called {\it anomalous cyclotron resonance}.
The peak height is relatively larger in the topological case than in the non-topological case,
and we conclude that existence of the anomalous peaks is a universal property of Kondo insulators
with band inversion.
Furthermore, we will see later that the anomalous cyclotron resonance shares common features 
in both non-topological and topological cases.
Such anomalous low energy peaks are absent in non-band inverted insulators, but are made possible
because of the characteristic Landau level structures in the Kondo insulators.
As seen in Fig.~\ref{fig:energy}, there are Landau levels densely distributed 
near the edges of the lowest positive energy band
($\varepsilon_{nm}>\mu$, shown in red) and the highest negative energy band ($\varepsilon_{nm}<\mu$, shown 
in blue), corresponding to the previously mentioned crossover in Landau level's character.
Transitions between the $n\leftrightarrow n+1$ Landau levels within each band edge in the same color
are allowed at finite temperature,
resulting in the low energy peaks in $\sigma(\omega)$,
while such band edge Landau levels are absent in a non-band inverted insulator.
Therefore, these anomalous peaks are a direct consequence of the unusual Landau level structure,
essentially sharing the common origin with quantum oscillations within the present model.

An observation of these peaks will strongly support the mechanism for quantum oscillations proposed in 
the previous theories~\cite{KnolleCooper2015,Zhang2016,Pal2016,Ram2017,
Shen2018,Grubinskas2018,Peters2019},
and hence provide a clue for the origin of the experimentally observed quantum oscillations.
If the magnetic field is increased further, $b>b_c$, so that the system is metallic,
one can see that the low energy peaks are further enhanced,
corresponding to heavy fermion formation in the metallic regime.
For the Kondo insulators SmB$_6$ and YbB$_{12}$, the frequency of the light should be smaller than 
$1\sim2$ meV corresponding to their insulating gaps, and also in between $\omega_f$ and $\omega_c$ (they are $\omega_f\sim0.1$ meV, $\omega_c\sim1$ meV for example at $B=10$ T in the present model). 
Besides, a large magnetic field $B>10$ T would be required to observe the anomalous cyclotron resonance.

\begin{figure}[htbp]
\begin{tabular}{cc}
\begin{minipage}{0.5\hsize}
\begin{center}
\includegraphics[width=\hsize]{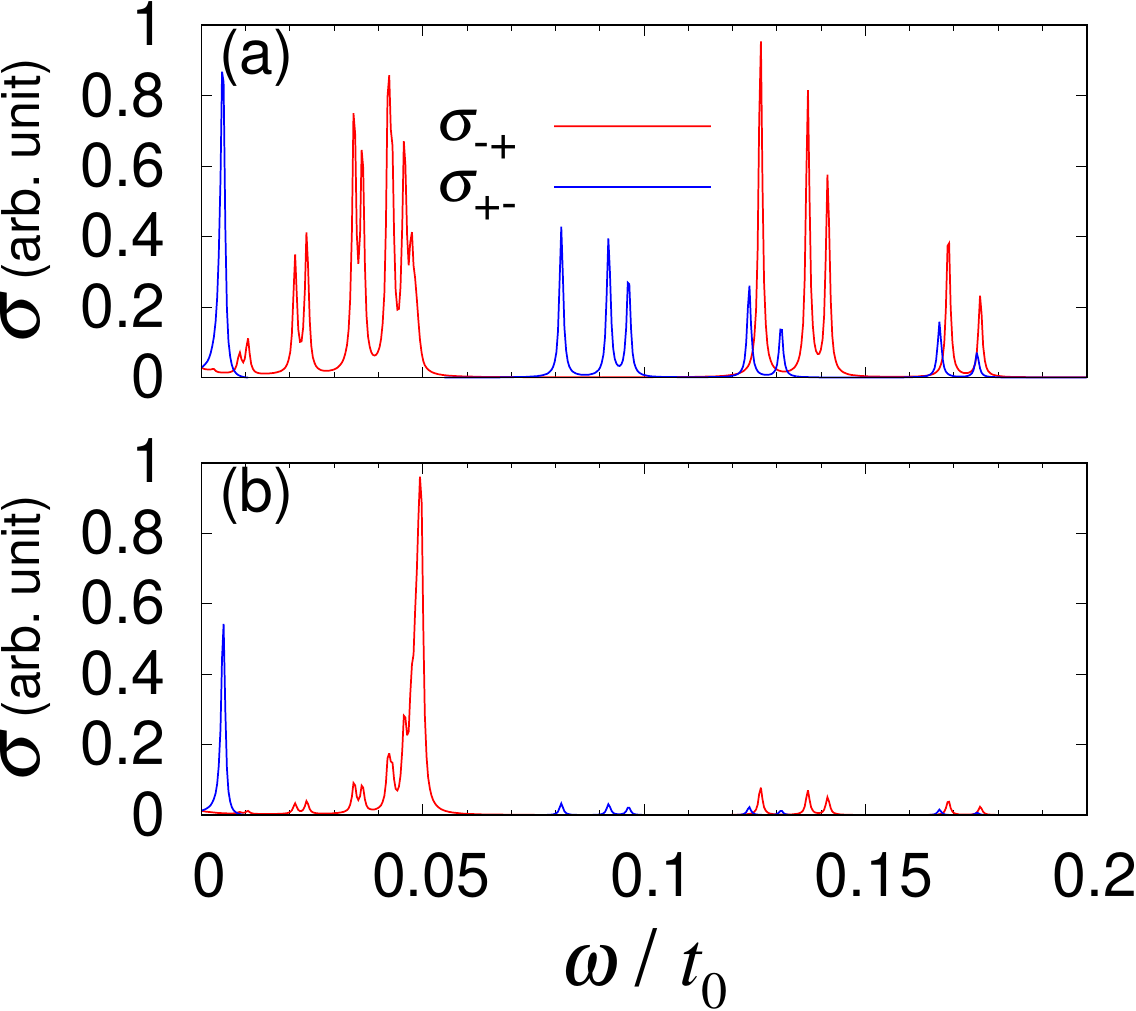}
\end{center}
\end{minipage}
\begin{minipage}{0.5\hsize}
\begin{center}
\includegraphics[width=\hsize]{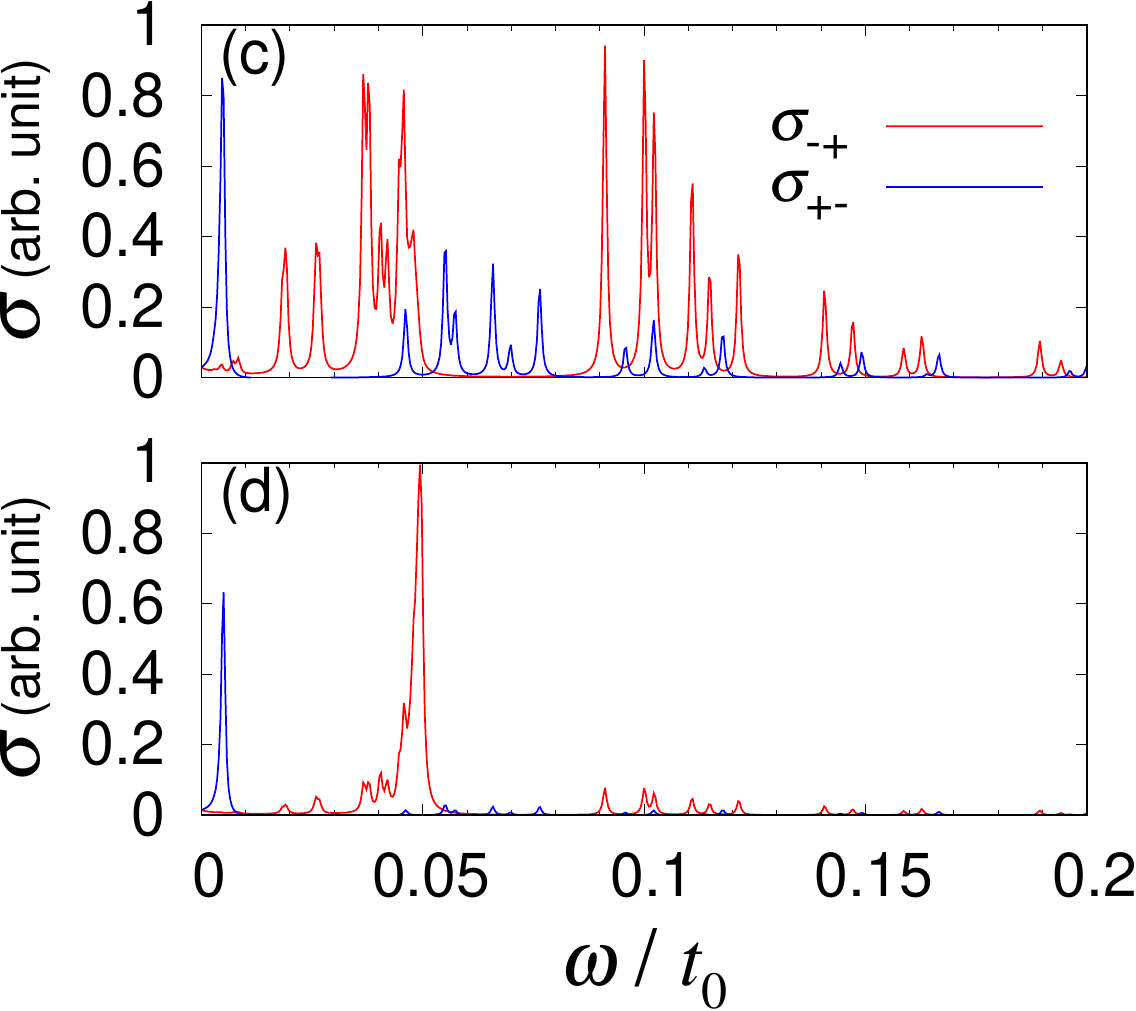}
\end{center}
\end{minipage}
\end{tabular}
\caption{Real part of the magneto-optical conductivity (in arbitrary unit) for 
(a), (b) the non-topological insulator  and 
(c), (d) the topological insulator at $b\simeq 0.83b_c$.
Temperature is (a) $T=0.05t_0=0.5\Delta_0$, (b) $T=0.2t_0=2\Delta_0$,
(c) $T=0.034t_0\simeq0.5\Delta_1$, and (d) $T=0.13t_0\simeq2\Delta_1$.
The red curve is Re$\sigma_{-+}(\omega)$ and the blue one is
Re$\sigma_{+-}(\omega)$. 
}
\label{fig:omT}
\end{figure}

Next, we discuss temperature dependence of the conductivity.
Figure \ref{fig:omT} shows $\sigma(\omega)$ for different temperatures 
at the magnetic field $b=0.05\simeq 0.83 b_c$.
At an intermediate temperature $T=0.5\Delta_{0,1}$,
the anomalous low energy peaks are now comparable
to the high energy peaks. 
Similarly to the $T=0.07\Delta_{0,1}$ case, these low energy peaks in $\sigma_{-+}(\omega)$ are located in 
the region $\omega_f=0.005t_0\leq \omega\leq \omega_c=0.05t_0$,
while there is only one low energy peak at $\omega\simeq\omega_f$ in $\sigma_{+-}(\omega)$.
The low energy peaks become sharper at high temperature, $T=2\Delta_{0,1}$,
and there is basically a single peak at $\omega=\omega_c$ in
$\sigma_{-+}(\omega)$ and at $\omega=\omega_f$ in $\sigma_{-+}(\omega)$.
These two peaks at high temperature simply correspond to the unhybridized conduction electrons and $f$-electrons,
which are directly seen in Eq.~\eqref{eq:sigma};
contributions from high energy states with large Landau level indices $n$ become dominant at very high temperature
since their momenta $\pi_{\pm}\sim \sqrt{n}/l_B$ are large,
and hence one can neglect hybridization matrix elements $v_0, v_1a\sqrt{n}/l_B\ll \omega_cn,\omega_fn$ in
the Hamiltonian ${\mathcal H}_n$. 
Then a straightforward calculation leads to 
$\sigma_{-+}(\omega\geq0)\propto \omega_c/(m_c^2T)\delta(\omega-\omega_c)$
and $\sigma_{+-}(\omega\geq0)\propto\omega_f/(m_f^2T)\delta(\omega-\omega_f)$
in the limit $\gamma\rightarrow0$.
The conduction and $f$-electrons  form a hybridized state at low temperature
for which energy levels are non-linear in the magnetic field $B$ and 
located in between the unhybridized conduction electron energy and $f$-electron energy.
This evolution of the system is a (non-interacting variant of) Kondo effect
and it would be useful for studying evolution of temperature dependent electronic structures 
under magnetic fields where photoemission experiments~\cite{smb6_photo_review,Okawa2015} are generally difficult.

\begin{figure}[htbp]
\begin{tabular}{cc}
\begin{minipage}{0.5\hsize}
\begin{center}
\includegraphics[width=\hsize]{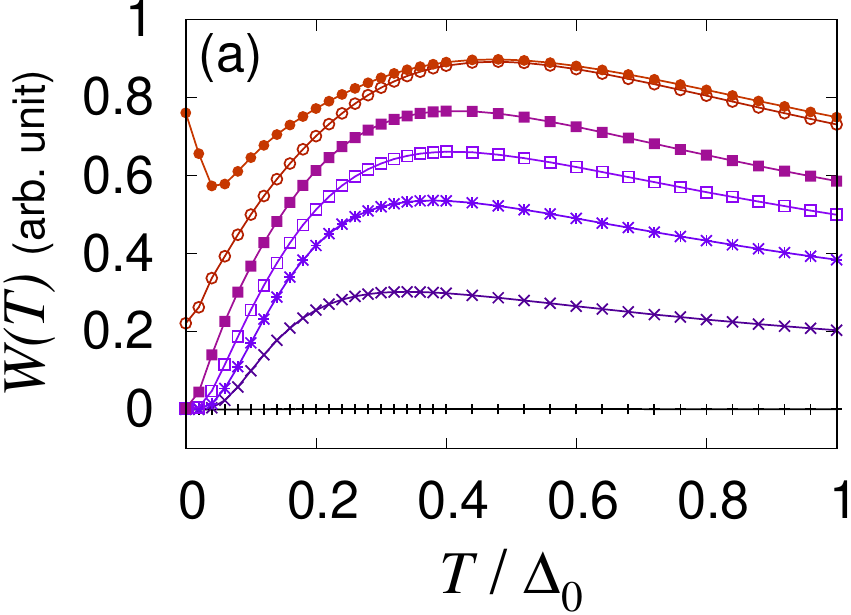}
\end{center}
\end{minipage}
\begin{minipage}{0.5\hsize}
\begin{center}
\includegraphics[width=\hsize]{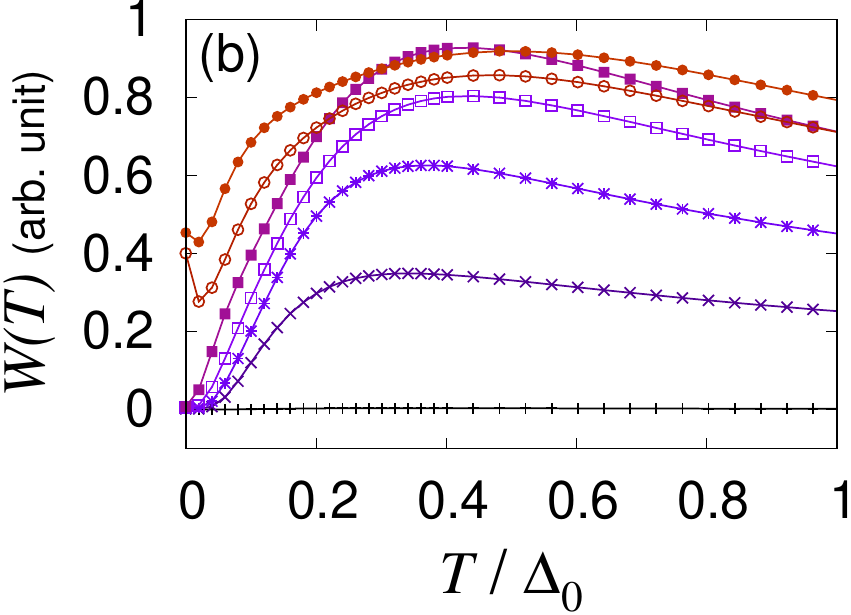}
\end{center}
\end{minipage}
\end{tabular}
\caption{Temperature dependence of the low energy spectral weight $W_{-+}$ in
(a) the non-topological and (b) topological Kondo insulators.
The magnetic fields are $b=0.01,0.02,0.03,0.04,0.05,0.06,0.07$ from the bottom (black) to top (dark-orange).
$W_{-+}$ for $b=0.01$ is much smaller than those for $b\geq0.02$.
}
\label{fig:W}
\end{figure}

The temperature dependence is well characterized by 
low energy spectral weight of the conductivity,
\begin{align}
W_{-+}=\int_{\Omega_1}^{\Omega_2}d\omega \textrm{Re}\sigma_{-+}(\omega).
\end{align}
We choose the cut off energy scales $\Omega_{1,2}$ so that only the anomalous
low energy peaks are taken into account; 
$\Omega_1=0$ and $\Omega_2=\omega_c-10\gamma$ for a given magnetic field.
Although the precise value of $W_{-+}$ depends on a choice of $\Omega_{1,2}$,
contributions from unhybridized conduction and $f$-electrons are suppressed with the present $\Omega_{1,2}$.
Besides,
only the transitions between states at Landau levels $n\leftrightarrow n+1$ with energies
$\Omega_1\lesssim |\varepsilon_{n+1,m}-\varepsilon_{nm'}\pm\gamma|\lesssim \Omega_2$ contribute to $W_{-+}$.
Figure \ref{fig:W} shows $W_{-+}$ for several values of magnetic fields $b$.
We see that $W_{-+}$ develops as temperature is reduced via the Kondo effect and it becomes
maximum roughly around $T\sim 0.3\Delta_{0,1}\textrm{ - }0.4\Delta_{0,1}$.
At low temperature,
the weight behaves as $W_{-+}(T)\sim \exp(-\Delta_{-+}/T)$ with a gap $\Delta_{-+}$ 
in the insulating regime.
$\Delta_{-+}$ corresponds to the minimum value of $|\varepsilon_{nm}-\mu|$ for the given 
magnetic field, because the temperature factor of $W_{-+}$ behaves as
$|f(\varepsilon_{n+1,m})-f(\varepsilon_{nm'})|
\simeq e^{-|\varepsilon_{n+1,m}-\mu|/T}(1-e^{-|\varepsilon_{n+1,m}-\varepsilon_{nm'}|/T})$ e.g. for 
$|\varepsilon_{n+1,m}-\mu|<|\varepsilon_{nm'}-\mu|$.
$W_{-+}$ is enhanced as the magnetic field increases in both non-topological and topological
Kondo insulators in a similar fashion, which again means that the enhancement is a universal property of
the Kondo insulators with band inversion.
In this way, the low energy peaks can become pronounced even in the insulating regime when 
the system is close to the insulator-metal transition point.
The gap $\Delta_{-+}$ becomes zero at the critical point $b=b_c\simeq 0.058$.
For larger magnetic fields $b=0.06, 0.07>b_c$ where the system is metallic,
there exist resonance peaks in $\sigma_{-+}(\omega)$ in between $\omega_f\leq \omega\leq \omega_c$ 
corresponding to heavy fermions and
$W_{-+}$ remains non-zero at $T=0$.

Our results for the clean non-interacting model 
imply that the anomalous low energy peaks corresponding to the characteristic Landau level structure
are most clearly visible around an intermediate temperature regime
($T\sim 0.3\Delta_{0,1}\textrm{ - }0.4\Delta_{0,1}$ in the above calculation).
In real materials, however, sharp peaks will be smeared by disorder and interactions especially at temperature above Kondo temperature which is roughly $T_K\sim \Delta_{0,1}$
~\cite{Appel1978,Takada1978,Ando1982,Asano1998,Kanki1997}, and
sufficiently low temperature would be required for an experiment.
At the same time, signal amplitudes would be reduced at low temperature, which
is a common subtle issue also for observables where only states near the chemical 
potential contribute~\cite{Pal2016,hetero1}.
Nevertheless, quantum oscillations in static longitudinal transport have indeed been observed not only in YbB$_{12}$
but also in band inverted insulating quantum wells at low temperature~\cite{Xiang2018,hetero2,hetero3},
and discussed theoretically~\cite{Peters2019}.
Since cyclotron resonance can be observed at relatively high temperature in general,
anomalous cyclotron resonance can be seen in some intermediate temperature range
below Kondo temperature and provide a useful insight for the observed quantum oscillations.

Finally, we qualitatively discuss effects of three dimensionality for which the response function
is $\sigma(\omega,q_z)$ for a light perpendicular to the $xy$-plane with an in-plane polarization.
Since we are interested in the frequency range below the activation gap $\Delta_{0,1}=O$(meV) 
or even smaller due to the renormalization effect,
wavelength of the light used in a cyclotron resonance experiment will be much longer than a
quasi-particle  mean free path.
Such a long wavelength may be regarded as the $q_z\rightarrow0$ limit and a resonance peak
position at $\omega=(\varepsilon_{n+1,m}(k_z+q_z)-\varepsilon_{nm'}(k_z))$ 
remains almost unchanged from that for a two-dimensional
system.
Therefore, the qualitative behaviors discussed in the present study will still hold for three dimensions.
In case of a topological Kondo insulator, 
there are Zeeman split surface states and they may give additional contributions to cyclotron resonance,
but they would be subdominant for a thick sample.
In a thin film, however, the surface mode contribution could be significant. 
A cyclotron resonance experiment with tilted field angles may provide useful information about the dimensionality of the carriers for the quantum oscillations.
There is another additional contribution even in two-dimensions arising from electron spins,
but it could be separated from cyclotron resonance 
because its magnitude is usually small and the characteristic energy scale 
is Zeeman energy which differs from a cyclotron frequency especially in
an $f$-electron system.
The spin response under a magnetic field itself will contain important information on 
elecronic structures similarly to those at zero magnetic field
~\cite{spin1,spin2}.

\section{Summary}
We have investigated cyclotron resonance in Kondo insulators. 
We found that there exist anomalous low energy resonance peaks in the magneto-optical conductivity
well below the activation gap
even at low temperature.
The low energy peaks are enhanced near the insulator-metal phase transition driven by
a strong magnetic field.
This is a direct consequence of the characteristic Landau level structure and is a
dynamical counterpart of quantum oscillations in static quantities,
and also shares the same origin (i.e. crossover of the Landau level character).
At temperatures sufficiently higher than the activation gap,
there are essentially two peaks corresponding to the unhybridized conduction and $f$-electrons,
and these peaks develop to form the anomalous peaks at low temperature
via the Kondo effect.
Our results are qualitatively applicable not only to Kondo insulators
but also to other band inverted insulators such as 
a quantum well 
~\cite{hetero1,hetero2,hetero3}.

\section*{Acknowldgement}

We thank M. Oshikawa, Y. H. Matsuda, and R. Peters for valuable discussions.
This work
was supported by Grants-in-Aid for Scientific Research No.
JP17K14333 and KAKENHI on Innovative Areas ``J-Physics''
(No. JP18H04318).

{\it Note added.}---
After the submission of the present manuscript,
we became aware of a recent theoretical work on cyclotron resonance in correlated insulators
~\cite{note0}.
We also found experimental reports of cyclotron resonance in InAs/GaSb heterostructures
which discuss effects of the electron-hole hybridization
~\cite{note1,note2,note3,note4}.


%

\end{document}